\documentclass[prb,showpacs,preprintnumbers,twocolumn,amsmath,amssymb]{revtex4}
\usepackage{graphicx}
\usepackage{dcolumn}
\usepackage{bm}

\newcommand{\PrOsSb}{PrOs$_4$Sb$_{12}$}
\newcommand{\PrRuSb}{PrRu$_4$Sb$_{12}$}
\newcommand{\PrOsRuSb}{Pr(Os$_{1-x}$Ru$_x$)$_4$Sb$_{12}$}
\newcommand{\LaFeP}{LaFe$_4$P$_{12}$}
\newcommand{\LaOsSb}{LaOs$_4$Sb$_{12}$}
\newcommand{\LaRuSb}{LaRu$_4$Sb$_{12}$}

\begin{document}

\title{Superconductivity and crystalline electric field effects
in the filled skutterudite series \PrOsRuSb}

\author{N. A. Frederick}

\author{T. D. Do}

\author{P.-C. Ho}

\author{N. P. Butch}

\author{V. S. Zapf}

\author{M. B. Maple}

\affiliation{%
Department of Physics and Institute for Pure and Applied Physical
Sciences, University of California at San Diego, La Jolla, CA
}%

\date{\today}

\begin{abstract}
X-ray powder diffraction, magnetic susceptibility $\chi(T)$, and
electrical resistivity $\rho(T)$ measurements were made on single
crystals of the filled skutterudite series \PrOsRuSb.  One end of
the series ($x = 0$) is a heavy fermion superconductor with a
superconducting critical temperature $T_{c} = 1.85$ K, while the
other end ($x = 1$) is a conventional superconductor with $T_{c}
\approx 1$ K. The lattice constant $a$ decreases approximately
linearly with increasing Ru concentration $x$. As Ru (Os) is
substituted for Os (Ru), $T_{c}$ decreases nearly linearly with
substituent concentration and exhibits a minimum with a value of
$T_{c} = 0.75$ K at $x = 0.6$, suggesting that the two types of
superconductivity compete with one another. Crystalline electric
field (CEF) effects in $\chi_\mathrm{dc}(T)$ and $\rho(T)$ due to
the splitting of the Pr$^{3+}$ nine-fold degenerate Hund's rule $J
= 4$ multiplet are observed throughout the series, with the
splitting between the ground state and the first excited state
increasing monotonically as $x$ increases. The fits to the
$\chi_\mathrm{dc}(T)$ and $\rho(T)$ data are consistent with a
$\Gamma_{3}$ doublet ground state for all values of x, although
reasonable fits can be obtained for a $\Gamma_{1}$ ground state
for $x$ values near the end member compounds ($x = 0$ or $x = 1$).
\end{abstract}



\maketitle

\section{Introduction}

The filled skutterudite compound \PrOsSb{} was recently discovered
to be the first Pr-based heavy fermion superconductor, with a
superconducting transition temperature $T_{c} = 1.85$ K and an
effective mass $m^* \approx 50~m_{e}$, where $m_{e}$ is the free
electron mass.\cite{Bauer02a,Maple02a} Features in the dc magnetic
susceptibility $\chi_\mathrm{dc}(T)$, specific heat $C(T)$,
electrical resistivity $\rho(T)$, and inelastic neutron scattering
(INS) can be associated with the thermally dependent population of
the ninefold degenerate Pr$^{3+}\ J = 4$ Hund's rule multiplet
split by a cubic crystalline electric field (CEF). These data
suggest that the ground state of \PrOsSb{} is a $\Gamma_{3}$
doublet, separated from a $\Gamma_{5}$ triplet first excited state
by $\sim 10$ K.\cite{Bauer02a,Maple02a}  The possibility of a
$\Gamma_{1}$ singlet ground state has also been put forward based
on other measurements,\cite{Aoki02,Kohgi03} some of which also
consider tetrahedral symmetry operators in their calculations of
the CEF Hamiltonian of \PrOsSb.\cite{Takegahara01} It has been
proposed that the superconductivity in \PrOsSb{} may be due to
quadrupolar fluctuations,\cite{Bauer02a} a claim that has been
supported by $\mu$SR\cite{MacLaughlin02} and
Sb-NQR\cite{Kotegawa03} measurements, which indicate a
strong-coupling isotropic energy gap of $2\Delta \approx
5k_{B}T_{c}$. Other intriguing effects are seen in \PrOsSb,
including multiple superconducting
transitions\cite{Maple03,Vollmer03,Oeschler03a} and
phases,\cite{Izawa03} and an ordered phase that is observed in
high magnetic fields and low temperatures.\cite{Ho02}  This high
field ordered phase (HFOP), which is seen in measurements of
$\rho(T)$,\cite{Ho02} $C(T)$,\cite{Vollmer03,Aoki02} magnetization
$M(T)$,\cite{Ho03} thermal expansion
$\alpha(T)$,\cite{Oeschler03a} and magnetostriction
$\lambda(T)$\cite{Oeschler03b} in a magnetic field $H$, as well as
measurements of $\rho(H)$ isotherms,\cite{Ho03} appears to be
related to the crossing of the CEF energy levels in magnetic
fields.\cite{Vollmer03,Frederick03a} In addition, neutron
diffraction experiments\cite{Kohgi03} indicate the presence of
quadrupolar effects in the HFOP, analagous to those seen in
PrPb$_{3}$.\cite{Tayama01} \PrOsSb{} has proven to be a unique
compound, and will continue to provide a fertile area of research
for many years.

The isostructural compound \PrRuSb{} displays superconductivity
below $T_{c} \approx 1.0$ K and possesses an electronic specific
heat coefficient $\gamma$ approximately $5$ to $10$ times smaller
than \PrOsSb, identifying it at as a conventional metal, or at
most a borderline heavy fermion metal.\cite{Takeda00} It was
previously reported, based on measurements of
$\chi_\mathrm{dc}(T)$, to possess a $\Gamma_{1}$ ground state and
a $\Gamma_{4}$ triplet first excited state $\approx 70$ K above
the ground state.\cite{Takeda00}  A later measurement of $\rho(T)$
also supported this CEF level scheme.\cite{Abe02}  \PrRuSb{}
appears to be a BCS-like weak-coupling superconductor, with an
isotropic s-wave energy gap of $2\Delta \approx 3k_{B}T_{c}$, as
determined from Sb-NQR measurements.\cite{Yogi03} At the present
time, no quadrupolar effects or features resembling the HFOP seen
in \PrOsSb{} have been reported in \PrRuSb.

The substitution of \PrRuSb{} into \PrOsSb{} to form \PrOsRuSb{}
was undertaken to investigate the evolution of the
superconductivity, the CEF energy level scheme, and the heavy
fermion state with Ru-doping, and to investigate the relationship,
if any, between these three phenomena. The present study focuses
on measurements of $\chi(x,T)$ and $\rho(x,T)$, which have
revealed the $x$-dependencies of $T_{c}$ and the splitting between
the CEF ground state and the first excited state.  We are also in
the process of investigating the heavy fermion state via
measurements of $C(T)$, and the upper critical field $H_{c2}(T)$
through measurements of $\rho(T,H)$ (which will also reveal the
$x$-dependence of the HFOP), and will report these results in a
future publication.

\section{Experimental Details}

Single crystals of \PrOsRuSb{} were grown using an Sb flux method.
The elements (Ames $5$N Pr, Colonial Metals $3.5$N Os and $3$N Ru,
and Alfa Aesar $6$N Sb) were sealed under $150$ Torr Ar in a
carbon-coated quartz tube in the ratio $1:4-4x:4x:20$, heated to
$1050~^{\circ}$C at $50~^{\circ}$C/hr, then cooled at
$2~^{\circ}$C/hr to $700~^{\circ}$C. The samples were then removed
from the furnace and the excess Sb was spun off in a centrifuge.
The crystals were removed from the leftover flux by etching with
dilute Aqua Regia (HCl:HNO$_3$:H$_{2}$O $= 1:1:3$).

\begin{figure}[tbp]
\begin{center}
\includegraphics[angle=270,width=3.4in]{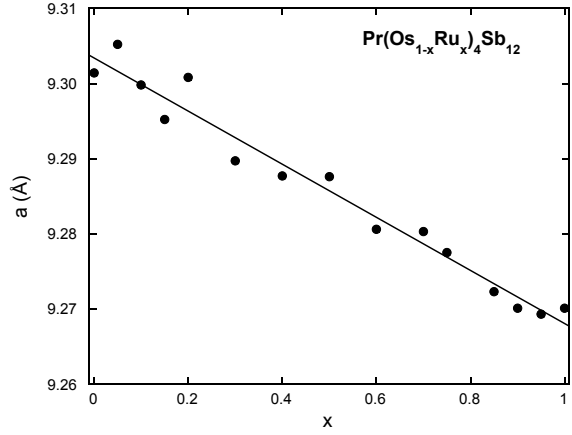}
\end{center}
\caption{Lattice parameter $a$ as a function of Ru concentration
$x$.  The solid line is a linear least squares fit to $a$ vs $x$.}
\label{xray}
\end{figure}

X-ray powder diffraction measurements were made at room
temperature using a Rigaku D/MAX B x-ray machine.  The only
significant impurities in any of the samples were identified with
free Sb that was still attached to the crystals.  Each \PrOsRuSb{}
sample crystallized in the \LaFeP{} structure \cite{Braun80} with
a lattice constant $a$ that decreased roughly linearly with
increasing Ru concentration $x$, as displayed in Fig.\ \ref{xray}.
A silicon standard was used in order to achieve a more accurate
determination of the lattice constant.  Measurements of
$\chi_\mathrm{dc}$ vs temperature $T$ were made in a magnetic
field $H$ of $0.5$ tesla between $1.8$ and $300$ K in a commercial
Quantum Design superconducting quantum interference device (SQUID)
magnetometer. Measurements of $\rho$ and $\chi_\mathrm{ac}$ were
made as a function of $T$ down to $1.2$ K in a $^4$He cryostat
and, for several of the samples, down to $0.1$ K in a
$^3$He-$^4$He dilution refrigerator.

\section{Results}

\subsection{Magnetic Susceptibility}

\begin{figure}[tbp]
\begin{center}
\includegraphics[angle=270,width=3.4in]{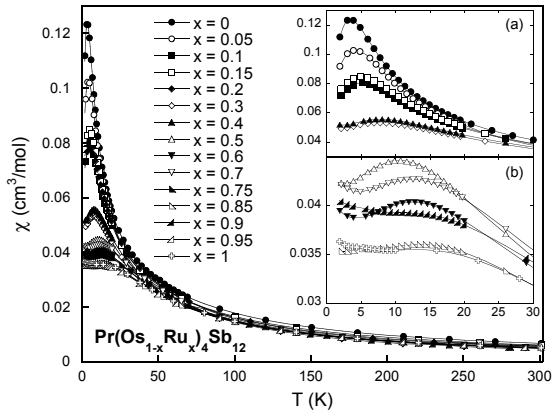}
\end{center}
\caption{dc magnetic susceptibility $\chi_\mathrm{dc}$ as a
function of temperature $T$ between $1.8$ and $300$ K for single
crystals of \PrOsRuSb. Inset (a): $\chi_\mathrm{dc}$ vs $T$
between $1.8$ and $30$ K, showing the evolution of the peak due to
crystalline electric field effects for the Ru concentrations $x =
0$ to $x = 0.4$.  The $x = 0.2$ sample has been removed for
clarity.  Inset (b): as inset (a), but for Ru concentrations $x =
0.5$ to $x = 1$. The samples with $x = 0.75$ and $x = 0.95$ were
removed for clarity.} \label{chidc}
\end{figure}

Displayed in the main portion of Fig.\ \ref{chidc} is a plot of
the dc magnetic susceptibility $\chi_\mathrm{dc}$ as a function of
temperature $T$ for single crystals of \PrOsRuSb{} with various
values of $x$. Above $T \approx 100$ K, the inverse magnetic
susceptibility $1/\chi_\mathrm{dc}$ is linear, indicating
Curie-Weiss behavior. The data have been corrected for excess Sb
by assuming that the high temperature effective moment,
$\mu_\mathrm{eff}$, of Pr should be equal to the Hund's rule free
ion value of $3.58~\mu_{B}$ for Pr$^{3+}$, where $\mu_{B}$ is the
Bohr magneton. Any deviation from this value was attributed to
free Sb inclusions in the \PrOsRuSb{} crystals. The most
significant effect on $\chi_\mathrm{dc}$ from this correction was
not the small diamagnetic Sb signal but instead the change in
overall scaling due to the difference in mass used to calculate
$\chi_\mathrm{dc}$ in units of cm$^3$/mol from the raw
magnetization data.  The calculated percentages of mass attributed
to Sb out of the total sample volume for all values of $x$ are
listed in Table \ref{results}.  The estimated value of the Sb mass
depends slightly on the CEF ground state used to make the fit
correction; only the values for a $\Gamma_{3}$ ground state are
given for simplicity.

All of the \PrOsRuSb{} samples exhibit features (peaks or
plateaus) in $\chi_\mathrm{dc}$ that can be attributed to CEF
effects. These features are the focus of the two insets in Fig.\
\ref{chidc}. The low temperature $\chi_\mathrm{dc}$ data for the
samples from $x = 0$ to $x = 0.4$ are shown in Fig.\
\ref{chidc}(a), while Fig.\ \ref{chidc}(b) similarly displays data
for the samples from $x = 0.5$ to $x = 1$.  An explanation of the
fits used to determine the CEF parameters from the
$\chi_\mathrm{dc}$ data, as well as the parameters themselves, is
given in section \ref{CEF}.

\begin{figure}[tbp]
\begin{center}
\includegraphics[angle=270,width=3.4in]{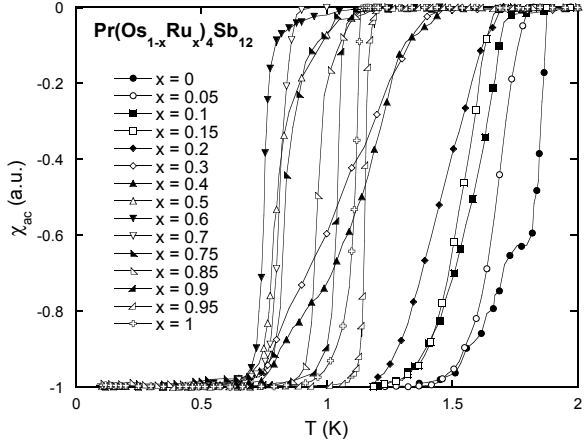}
\end{center}
\caption{ac magnetic susceptibility $\chi_\mathrm{ac}$ as a
function of temperature $T$ between $0.1$ and $2$ K for single
crystals of \PrOsRuSb. The data have been normalized to $0$ at $T
= 2$ K and to $-1$ at $T = 0$ K for clarity.} \label{chiac}
\end{figure}

Low temperature ($< 2$ K) ac magnetic susceptibility
$\chi_\mathrm{ac}$ vs $T$ data for \PrOsRuSb{} are shown in Fig.\
\ref{chiac}.  A sharp diamagnetic transition can be seen for all
values of $x$, indicating the presence of superconductivity. The
superconducting critical temperature $T_{c}$ for each
concentration was determined from the data displayed in Fig.\
\ref{chiac} as the midpoint of the diamagnetic transition.  A plot
of $T_{c}$ vs $x$ is displayed in Fig.\ \ref{Tc}, and is discussed
further in section \ref{SC}. An additional feature of note is the
step-like structure that appears in the $\chi_\mathrm{ac}$ data
for \PrOsSb. Since double superconducting transitions have been
observed in specific heat and thermal expansion measurements on
both collections of single crystals and individual single
crystals,\cite{Maple03,Vollmer03,Oeschler03a} it is reasonable to
assume that this step in the diamagnetic transition for \PrOsSb{}
is also due to an intrinsic second superconducting phase instead
of a variation of $T_{c}$ throughout the multiple crystals used in
the $\chi_\mathrm{ac}$ measurements. None of the other
concentrations display significant structure in their
superconducting transitions, although the transitions for $x =
0.3$ and $x = 0.4$ are much wider than for the other
concentrations.  This may be due to a variation of $T_{c}$ between
individual crystals for these two concentrations.

\subsection{Electrical Resistivity}

\begin{figure}[tbp]
\begin{center}
\includegraphics[angle=270,width=3.4in]{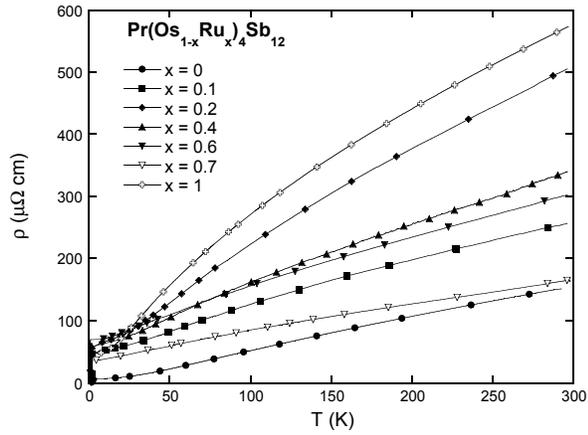}
\end{center}
\caption{Electrical resistivity $\rho$ as a function of
temperature $T$ between $0.4$ and $300$ K for single crystals of
\PrOsRuSb{} with various values of $x$ between $0$ and $1$. The
samples with $x = 0.05$, $x = 0.15$, and $x = 0.9$ were removed
for clarity.} \label{rho}
\end{figure}

Fig.\ \ref{rho} displays high-temperature electrical resistivity
$\rho$ vs $T$ data for \PrOsRuSb{} for various values of $x$
between $0$ and $1$. The values of $\rho$ at room temperature,
$\rho(300\ \mathrm{K})$, and the extrapolated values of $\rho$ at
zero temperature from fits to the $\rho(T)$ data based on
calculations of $\rho(T)$ that incorporate CEF splitting of the
Pr$^{3+}\ J = 4$ multiplet (see section \ref{CEF}), $\rho(0\
\mathrm{K})$, are listed in Table \ref{results}.  Also listed in
Table \ref{results} is the residual resistivity ratio (RRR),
defined as $\rho(300\ \mathrm{K})/\rho(0\ \mathrm{K})$.  It is
surprising that the RRR of \PrRuSb{} is so much lower than that of
\PrOsSb{}, since they are both stoichiometric compounds and would
be expected to have a low residual resistivity. A previous
measurement of \PrRuSb{} found $\rho(300\ \mathrm{K}) = 632\
\mu\Omega$ cm and an RRR of $25$,\cite{Takeda00} in reasonable
agreement with the data presented in this paper.  The low RRR of
\PrRuSb{} is not presently understood.

\begin{figure}[tbp]
\begin{center}
\includegraphics[angle=270,width=3.4in]{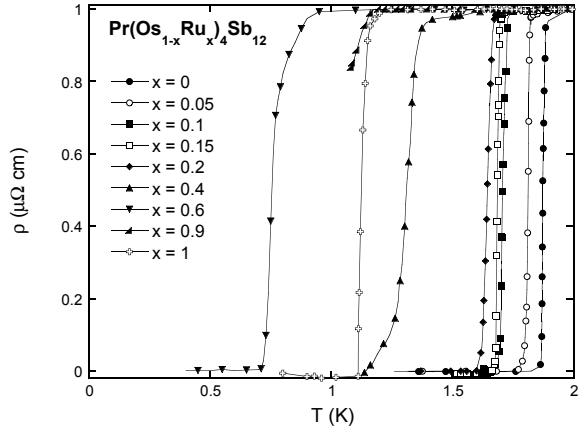}
\end{center}
\caption{Electrical resistivity $\rho$ as a function of
temperature $T$ between $0.4$ and $2$ K for single crystals of
\PrOsRuSb{} with various values of $x$ between $0$ and $1$,
normalized to their values at $2$ K.  The data for the sample with
$x = 0.7$ is not shown because it did not superconduct down to the
lowest temperature measured (see text for details). Similarly, the
superconducting transition for $x = 0.9$ is not complete due to
the limits of the experiment.} \label{rhoSC}
\end{figure}

The electrical resistivity of \PrOsRuSb{} below $T = 2$ K is shown
in Fig.\ \ref{rhoSC}. The data have been normalized to their
values at $2$ K in order to emphasize the superconducting
transitions. The $x = 0.7$ sample did not display the onset of
superconductivity down to the lowest measured temperatures and no
data for this sample are shown in this plot; the heating due to
large contact resistances in the $x = 0.7$ and $x = 0.9$ samples
precluded measurements below $1$ K. The superconducting
transitions as determined from $\rho(T)$ are in reasonable
agreement with those measured inductively (Fig.\ \ref{chiac}), and
the plot of $T_{c}$ vs $x$ is discussed in the following section.

\section{Analysis and Discussion}

\subsection{Superconductivity}
\label{SC}

\begin{figure}[tbp]
\begin{center}
\includegraphics[angle=270,width=3.4in]{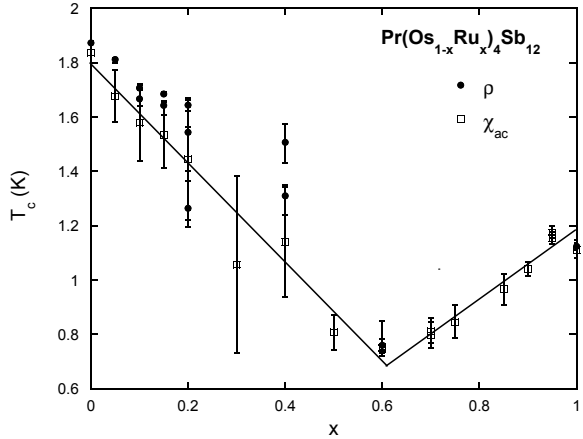}
\end{center}
\caption{Superconducting critical temperature $T_{c}$ vs Ru
concentration $x$ for \PrOsRuSb.  Filled circles: $T_{c}$
extracted from electrical resistivity $\rho$.  Open squares:
$T_{c}$ determined from ac magnetic susceptibility
$\chi_\mathrm{ac}$. The straight lines are guides to the eye.}
\label{Tc}
\end{figure}

The dependence of the superconducting transition temperature
$T_{c}$ on Ru concentration $x$ for \PrOsRuSb{} is shown in Fig.\
\ref{Tc}.  Several concentrations have more than one data point
associated with them, which are from measurements of different
crystals.  These additional measurements were not shown in Figs.\
\ref{rho} and \ref{rhoSC} or listed in Table \ref{results} in the
interest of clarity.  The RRRs were nearly identical for all
crystals of a given concentration, with the exception of the $x =
0.2$ samples where the crystal with the lowest $T_{c}$ in Fig.\
\ref{Tc} had an RRR about half of that measured for the other two
$x = 0.2$ samples, one of which is listed in Table \ref{results}.
The vertical bars in Fig.\ \ref{Tc} are a measure of the width of
the superconducting transitions, taken to be the $10\%$ and $90\%$
values of the resistance change associated with the transition.

The trend of the $T_{c}$ vs $x$ data shown in Fig.\ \ref{Tc}
suggests a competition between the two different types of
superconductivity seen in \PrOsSb{} and \PrRuSb.  This competition
suppresses $T_{c}$ from both ends, culminating in a minimum of
$T_{c} = 0.75$ K near $x = 0.6$.  Specific heat measurements are
in progress, and it will be interesting to see if the heavy
fermion state can be correlated with $T_{c}$.  The persistence of
superconductivity throughout the series is unusual, as for heavy
fermion f-electron superconductors both magnetic and nonmagnetic
impurities generally produce relatively rapid depressions of
$T_{c}$.  When the impurity is of an element that would produce an
isostructural superconducting compound, the trend is not as clear.
For example, the U$_{1-x}$La$_{x}$Pd$_{2}$Al$_{3}$ system is
similar to the \PrOsRuSb{} system in that one end member compound,
UPd$_{2}$Al$_{3}$, is a heavy fermion superconductor, while the
other end member compound, LaPd$_{2}$Al$_{3}$, is a conventional
BCS superconductor.  Unlike \PrOsRuSb, however, superconductivity
is destroyed upon substitution on either end of the
series.\cite{Zapf01b}  This persistence of superconductivity
throughout of the \PrOsRuSb{} system for all values of $x$ is
observed in the CeCo$_{1-x}$Ir$_{x}$In$_{5}$ series of compounds,
which is also superconducting for all values of
$x$.\cite{Pagliuso02} This system's similarities to \PrOsRuSb{}
end there, because both end member compounds (CeCoIn$_{5}$ and
CeIrIn$_{5}$) are heavy fermion superconductors in which the
superconductivity is believed to be magnetically mediated and to
possess nodes in the energy gap $\Delta({\bf
k})$.\cite{Movshovich02}

This nodal energy gap structure may be in contrast with \PrOsSb,
where $\mu$SR\cite{MacLaughlin02} and Sb-NQR\cite{Kotegawa03}
measurements indicate an isotropic energy gap, a condition which
could occur if the superconductivity in \PrOsSb{} was mediated by
quadrupolar fluctuations.  It is also generally the case that
superconductors with isotropic or nearly isotropic energy gaps are
relatively insensitive to the presence of nonmagnetic impurities.
Thus, the gradual decrease of $T_{c}$, and the presence of
superconductivity for all values of $x$ in \PrOsRuSb, provides
further evidence for an isotropic energy gap and quadrupolar
superconductivity in \PrOsSb, since \PrRuSb{} also possesses an
isotropic superconducting energy gap.\cite{Yogi03} The minimum in
$T_{c}$ near $x = 0.6$ could then be attributed to a shift from
quadrupolar mediated heavy fermion superconductivity to phonon
mediated BCS superconductivity.  On the other hand, thermal
conductivity measurements of \PrOsSb{} in a magnetic field have
been interpreted in terms of two distinct superconducting phases
in the $H-T$ plane, one with two point nodes in $\Delta({\bf k})$
in low fields, and another with six point nodes in $\Delta({\bf
k})$ at higher fields.\cite{Izawa03}  Since no thermal
conductivity measurements were reported for fields below $0.3$ T,
the structure of $\Delta({\bf k})$ is not known below this field.
It is conceivable that, just as a magnetic field induces a change
from a state with two point nodes into a state with six point
nodes, the state with two point nodes is itself induced from an
isotropic zero-field energy gap. Further measurements of the
energy gap symmetry in zero and low magnetic field could shed
light on this mystery.

\subsection{Crystalline Electric Field Effects}
\label{CEF}

\begin{figure}[tbp]
\begin{center}
\includegraphics[width=3.4in]{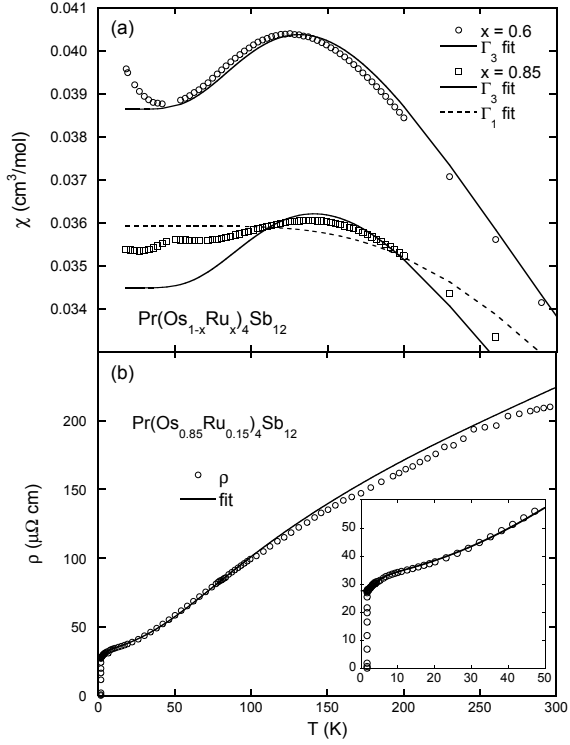}
\end{center}
\caption{Examples of CEF fits to the data.  (a): dc magnetic
susceptibility $\chi_\mathrm{dc}(T)$ for $x = 0.6$ and $x = 0.85$
for \PrOsRuSb.  The solid lines are fits assuming a $\Gamma_{3}$
doublet ground state and a $\Gamma_{5}$ triplet first excited
state, and the dashed line is a fit assuming a $\Gamma_{1}$
singlet ground state and a $\Gamma_{4}$ triplet first excited
state. (b): electrical resistivity $\rho(T)$ for $x = 0.15$
between $1$ K and $300$ K. The fit is for a $\Gamma_{3}$ ground
state and a $\Gamma_{5}$ first excited state.  Fits with a
$\Gamma_{1}$ ground state were qualitatively identical, and so are
not shown (see text for details).  Inset to (b): $\rho(T)$ for $x
= 0.15$ between $1$ K and $50$ K, displaying the low-temperature
curvature in greater detail.} \label{CEFexamples}
\end{figure}

\begin{figure}[tbp]
\begin{center}
\includegraphics[angle=270,width=3.4in]{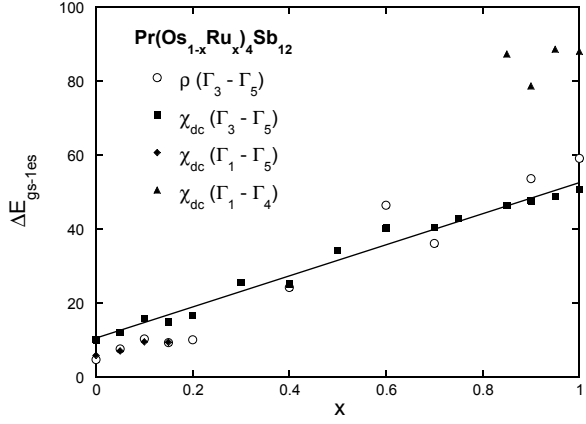}
\end{center}
\caption{The splitting between the ground state and first excited
state $\Delta E_\mathrm{gs-1es}$ vs Ru concentration $x$ for
\PrOsRuSb, calculated from fits of CEF equations to
$\chi_\mathrm{dc}(T)$ and $\rho(T)$, as described in the text. The
solid line is a linear fit to $\Delta E_\mathrm{gs-1es}$ for a
$\Gamma_{3}$ doublet ground state and a $\Gamma_{5}$ triplet first
excited state calculated from the $\chi_\mathrm{dc}(T)$ data. For
$x \leq 0.15$, a CEF energy level scheme with a $\Gamma_{1}$
singlet ground state and a $\Gamma_{5}$ first excited state also
provided a reasonable fit to the $\chi_\mathrm{dc}(T)$ data, while
a $\Gamma_{1}$ ground state with a $\Gamma_{4}$ triplet first
excited state was also a possible energy level scheme for $x \geq
0.85$.} \label{CEFvsx}
\end{figure}

The $\chi_\mathrm{dc}(T)$ and $\rho(T)$ data for \PrOsRuSb{} were
fit to equations including CEF effects, in a manner identical to
that reported previously.\cite{Bauer02a,Frederick03a}  The CEF
equations were derived from the Hamiltonian of Lea, Leask and Wolf
(LLW).\cite{Lea62} In the LLW formalism, the CEF energy levels are
given in terms of the parameters $x_\mathrm{LLW}$ and $W$, where
$x_\mathrm{LLW}$ is the ratio of the fourth and sixth order terms
of the angular momentum operators and $W$ is an overall energy
scale factor. It was assumed that the CEF parameter $y$ which
controls the tetrahedral $T_{h}$ crystalline symmetry contribution
to the Hamiltonian\cite{Takegahara01} was small; thus, the
calculations were made for a cubic $O_{h}$ crystalline symmetry.
Assuming that $y$ is small implies that the main contribution to
the crystalline electric field comes from the simple cubic
transition metal sublattice (Os or Ru), as opposed to the more
complicated tetrahedral Sb sublattice.  The $\chi_\mathrm{dc}(T)$
data for $x \leq 0.15$ could be reasonably fit with either a
$\Gamma_{3}$ or a $\Gamma_{1}$ ground state and a $\Gamma_{5}$
first excited state. As $x$ increases, the magnitude of the peak
in $\chi_\mathrm{dc}$ decreases more rapidly than the temperature
$T_\mathrm{max}$ at which the peak occurs. The peak also broadens
until it resembles a hump. These changes with $x$ make it
unreasonable to fit a $\Gamma_{1}$--$\Gamma_{5}$ CEF energy level
scheme to the higher $x$ data, since for these data an energy
level scheme with the correct $T_\mathrm{max}$ makes the peak too
sharp, while the correct hump shape results in a $T_\mathrm{max}$
that is too high. Thus, for the \PrOsRuSb{} samples with $x \geq
0.2$, a $\Gamma_{3}$ ground state best approximated the data. An
example of a fit with a $\Gamma_{3}$ ground state for $x = 0.6$ is
shown in Fig.\ \ref{CEFexamples}(a). A plot of the splitting
between the ground state and the first excited state vs $x$ is
shown in Fig.\ \ref{CEFvsx}, including all reasonable fits of the
$\chi_\mathrm{dc}(T)$ data.

The \PrOsRuSb{} samples with $x \geq 0.6$ all display upturns in
$\chi_\mathrm{dc}(T)$ at the lowest temperatures (inset (b) in
Fig.\ \ref{chidc}). If these upturns are due to the splitting of
the CEF energy levels in a small magnetic field $H$, then it is
expected that they would be more visible in the samples with large
$x$ (more Ru than Os), where $\chi_\mathrm{dc}$ is smaller at low
temperatures compared to the small $x$ (more Os than Ru) data. The
samples with $x \geq 0.75$, including \PrRuSb, also display
structure in these upturns that appear to be an additional peak
near $5$ K superimposed on the broad CEF hump, near the
temperature of the CEF peak in \PrOsSb. The smooth progression of
both the lattice parameter $a$ and $T_{c}$ indicates that there is
no macroscopic phase separation of \PrOsRuSb{} into \PrOsSb{} and
\PrRuSb. However, it is possible that the peak-like structure
could be due to inhomogeneous alloying of Os and Ru on an atomic
scale, wherein each Pr$^{3+}$ ion sees a distribution of Os or Ru
atoms, leading to a variation in the CEF throughout the crystal.
Unfortunately, this possibility would be difficult to establish in
the current experiments. The low-temperature upturn, especially in
\PrRuSb, could be attributed to either CEF splitting in $H$ or
paramagnetic impurities, both of which could produce a
low-temperature increase in $\chi_\mathrm{dc}$.

Takeda et al.\ reported that \PrRuSb{} had a $\Gamma_{1}$ singlet
ground state and a $\Gamma_{4}$ triplet first excited state, a CEF
configuration that exhibits a plateau in $\chi_\mathrm{dc}$ at low
temperatures.\cite{Takeda00} In the current experiment, the $x =
0.9$ and $x = 1$ samples are the only ones in which a plateau is
observed.  In addition, while the other samples with $x \geq 0.85$
have their peaks reasonably well described by a $\Gamma_{3}$
ground state, the fit predicts a saturation at $T = 0$ K that is
much lower than is observed in the data.  However, the low-$T$
upturn could be responsible for disguising both the maximum in $x
= 0.9$ and $x = 1$ and the low temperature saturation observed in
the other high Ru concentration samples. Accordingly, all the
\PrOsRuSb{} data with $x \geq 0.85$ were fit assuming both a
$\Gamma_{3}-\Gamma_{5}$ CEF energy level scheme and a
$\Gamma_{1}-\Gamma_{4}$ scheme, ignoring the low-temperature
upturn; the $x = 0.85$ fits are shown in Fig.\
\ref{CEFexamples}(b). Both fits are represented in the splitting
between the ground state and first excited state $\Delta
E_\mathrm{gs-1es}$ vs $x$ plot of Fig.\ \ref{CEFvsx}; the results
from all fits are also listed in Table \ref{results}.

The electrical resistivity $\rho(T)$ of \PrOsRuSb{} was fit by a
combination of scattering from impurities, the atomic lattice
(phonons), and temperature-dependent energy level populations due
to the CEF.\cite{Frederick03a}  The phonon contribution was
represented by the measured $\rho_\mathrm{lat}$ of \LaOsSb, an
isostructural reference compound without f-electrons, for all
values of $x$. This procedure was validated by reproducing the
results of Abe et al.\cite{Abe02} with \LaOsSb{} instead of
\LaRuSb; as expected, the $\rho_\mathrm{lat}$ data of the two
compounds appear to be nearly identical.  The CEF contribution to
$\rho(T)$ consists of two terms, representing magnetic exchange
and aspherical Coulomb scattering, which were assumed to be
equally important when fitting the data.\cite{Frederick03a} Just
as it was possible to fit $\rho(T)$ of \PrOsSb{} with either a
$\Gamma_{3}$ or a $\Gamma_{1}$ ground state, all of the
\PrOsRuSb{} data were indifferent to the choice of either ground
state.  The splitting between the ground state and the first
excited state (always a $\Gamma_{5}$ triplet) was also nearly
identical for fits with either ground state for a particular value
of $x$. In the interest of simplicity, for the $\rho(T)$ data,
only the splitting between $\Gamma_{3}$ and $\Gamma_{5}$, $\Delta
E_{3-5}$, is shown in Fig.\ \ref{CEFvsx}. The fit used to
calculate $\Delta E_{3-5}$ for $x = 0.15$ is shown in Fig.\
\ref{CEFexamples}(b). It is evident that $\rho(H)$ measurements
will be required to elucidate the CEF ground state from transport
measurements.\cite{Frederick03a}

It is unclear what effect the CEF ground state may have on the
superconductivity in \PrOsRuSb.  From a physical point of view, it
is reasonable that an abrupt change in the ground state would
produce an equally abrupt change in the physical properties.
However, it is difficult to conceive of a mechanism for this
occurrence in the context of the LLW theory, since it is based on
the interaction of the atomic lattice with a rare earth ion.  If
there is not an abrupt change in the lattice structure, one should
not expect an abrupt change in the CEF ground state. It is
therefore far more reasonable to consider a constant ground state,
with the excited state varying as the Ru substitution changes the
spacing of the atoms in the skutterudite lattice.  The present
data are most consistent with a constant $\Gamma_{3}$ ground
state, with the exception of the $x = 0.9$ and $x = 1$ data.
However, when $x$ is in the region $0.2 \leq x \leq 0.75$ a
$\Gamma_{3}-\Gamma_{5}$ CEF energy level scheme is the only one
which reasonably fits the $\chi_\mathrm{dc}(T, x)$ data. On the
other hand, the possibility cannot be ruled out that this deep in
a substituted system, a CEF analysis in the tradition of LLW may
be unreliable due to the distribution of the two substituents (Or
and Ru) in the near neighbor environment of each Pr$^{3+}$ ion.
The gradual metamorphosis of the $\chi_\mathrm{dc}$ data does
suggest that the CEF parameters are also changing gradually, but
this may be misleading. Further experiments as well as theoretical
analysis will be necessary to completely reveal the CEF ground
state and its relationship to the superconductivity.

\section{Summary}

The superconducting critical temperature $T_{c}$ and crystalline
electric field (CEF) parameters of single crystals of \PrOsRuSb{}
have been deduced through measurements of $\chi(T)$ and $\rho(T)$
for $0 \leq x \leq 1$ . The superconductivity, which is present
for all values of $x$, exhibits a change in the sign of the slope
in $T_{c}(x)$ near $x = 0.6$.  The CEF ground state may also
change from a $\Gamma_{3}$ ground state to a $\Gamma_{1}$ ground
state near this concentration, although more measurements are
necessary to confirm this possibility. It is possible that the
crossover from heavy fermion superconductivity that may be
mediated by quadrupolar interactions to nearly BCS
superconductivity occurs at this `pseudocritical' concentration
$x_{pc} = 0.6$.

\section*{Acknowledgements}

We would like to thank S. K. Kim and D. T. Walker for experimental
assistance, and E. D. Bauer for useful discussions. This research
was supported by the U.S. Department of Energy Grant
No.~DE-FG03-86ER-45230, the U.S. National Science Foundation Grant
No.~DMR-00-72125, and the NEDO International Joint Research
Program.


\newpage

\newpage

\begin{table}
\caption{Physical properties of \PrOsRuSb{} compounds. $x$ is the
concentration of Ru; $\rho(300\ \mathrm{K})$ is the electrical
resistivity $\rho$ at $300$ K; $\rho(0\ \mathrm{K})$ is $\rho$ at
$0$ K extrapolated from CEF fits (see text); RRR is the residual
resistivity ratio, defined as $\rho(300\ \mathrm{K})/\rho(0\
\mathrm{K})$; $\%$Sb is the percentage of the mass attributed to
free Sb in $\chi_\mathrm{dc}(T)$ assuming a $\Gamma_{3}$ ground
state; $x_\mathrm{LLW}$ and $W$ are the Lea, Leask and Wolf
parameters;\cite{Lea62} and $\Delta E_{a-b}$ is the energy
difference between ground state $\Gamma_{a}$ and first excited
state $\Gamma_{b}$.} \label{results}
\begin{tabular}{|c|ccc|r@{.}lccc|cccc|}
    \hline
 $x$ &  $\rho(300\ \mathrm{K})$ & $\rho(0\ \mathrm{K})$ & RRR &
 \multicolumn{2}{c}{$\%$Sb} & $x_\mathrm{LLW}$ & $W$ & $\Delta E_{3-5}$
 & $x_\mathrm{LLW}$ & $W$ & $\Delta E_{1-5}$ & $\Delta E_{1-4}$ \\
  & ($\mu\Omega$ cm) & ($\mu\Omega$ cm) &
  & \multicolumn{4}{c}{$\Gamma_{3}$ ground state} & (K)
  & \multicolumn{2}{c}{$\Gamma_{1}$ g.s.} & (K) & (K) \\
  \hline
0 & 155 & 1.67 & 93 & 25 & 0 & $-0.721$ & $-5.69$ & 10.1 & 0.500 & 1.99 & 5.87 & --- \\
0.05 & 235 & 18.7 & 13 & 15 & 1 & $-0.720$ & $-6.38$ & 12.1 & 0.484 & 1.47 & 7.08 & --- \\
0.1 & 259 & 46.0 & 5.6 & 21 & 3 & $-0.717$ & $-7.05$ & 15.9 & 0.462 & 1.31 & 9.54 & --- \\
0.15 & 215 & 27.0 & 8.0 & 15 & 6 & $-0.718$ & $-7.00$ & 14.9 & 0.452 & 1.11 & 9.43 & --- \\
0.2 & 510 & 54.0 & 9.4 & 27 & 3 & $-0.713$ & $-6.16$ & 16.6 & --- & --- & --- & --- \\
0.3 & --- & --- & --- & 8 & 0 & $-0.707$ & $-7.48$ & 25.5 & --- & --- & --- & --- \\
0.4 & 343 & 58.2 & 5.9 & 20 & 2 & $-0.702$ & $-6.43$ & 25.2 & --- & --- & --- & --- \\
0.5 & --- & --- & --- & 4 & 9 & $-0.687$ & $-6.06$ & 34.2 & --- & --- & --- & --- \\
0.6 & 305 & 67.4 & 4.5 & 6 & 6 & $-0.675$ & $-5.75$ & 40.3 & --- & --- & --- & --- \\
0.7 & 166 & 34.8 & 4.8 & 10 & 1 & $-0.663$ & $-4.81$ & 40.4 & --- & --- & --- & --- \\
0.75 & --- & --- & --- & 17 & 2 & $-0.669$ & $-5.54$ & 42.8 & --- & --- & --- & --- \\
0.85 & --- & --- & --- & 6 & 3 & $-0.670$ & $-6.05$ & 46.4 & $-0.737$ & 2.70 & --- & 87.4 \\
0.9 & 330 & 42.4 & 7.8 & 11 & 9 & $-0.646$ & $-4.58$ & 47.7 & $-0.872$ & 3.43 & --- & 78.8 \\
0.95 & --- & --- & --- & 20 & 8 & $-0.665$ & $-5.94$ & 48.8 & $-0.970$ & 5.51 & --- & 88.7 \\
1 & 578 & 41.8 & 14 & 7 & 4 & $-0.946$ & $-5.45$ & 50.8 & $-0.946$ & 4.95 & --- & 88.1 \\
 \hline
\end{tabular}
\end{table}

\end{document}